# Ultra cold neutron trap as a tool to search for dark matter with long-range radius of forces

A.P. Serebrov[*], O.M. Zherebtsov, A.K. Fomin, and M.S. Onegin

*St. Petersburg Nuclear Physics Institute, Gatchina, 188 300, St. Petersburg, Russia*

The problem of possible application of an ultracold neutron (UCN) trap as a detector of dark matter particles with long-range radius of forces has been considered. Transmission of small recoil energy in scattering is a characteristic of long-range forces. The main advantage of the ultracold neutron technique lies in possibility of detecting recoil energy as small as $10^{-7}$ eV. Here are presented constraints on the interaction potential parameters: $U(r) = a\ r^{-1} e^{-r/\lambda}$ for dark matter particles and neutrons, as well as on the density value of long-range dark matter on the Earth. The possible mechanism of accumulation of long-range dark matter on the Earth surface is considered and the factor of density increase on the Earth surface is evaluated. The results of the first experiment on search of astronomical day variation of ultracold neutron storage time are under discussion.

## I. INTRODUCTION

The problem of dark matter of the Universe is one of the most significant and exciting in modern physics [1]. The nature of dark matter is unknown though there are estimations of the quantity of dark matter in the Universe [2] and its location in the Galaxy. These estimations point to the fact that the dark matter density nearly 5 times exceeds [2] that of ordinary baryon matter. Dark matter is invisible in the observable spectral range but it manifests itself, for instance, through gravitational influence on movement of stars in galaxies and on movement of galaxies and clouds of hot gas in constellations of Galaxies [3], which confirms an obvious fact of its existence (q. v. [4]).The nature of dark matter is unknown, nevertheless, for example in [5, 6] heavy leptons are considered as candidates for the dark matter particles. We will deal with two statements to be reliably proved: 1) Interaction of dark matter with our matter is extremely weak, for instance, in case when WIMP-cross section particles $\leq 10^{-40}$ cm$^2$ are regarded as alternative of dark matter. This assumption is based on numerous experimental data on dark matter search [2], 2) In the Galaxy dark matter is localized in the so called halo. The latter is a widely recognized fact as the rate of movement of stars and clouds of cold gas in Galaxies with respect to the distance to the Galaxy center cannot be interpreted without involving additional invisible mass distributed in the Galaxy halo [2]. It is these observations which show that the Galaxy dark matter density is $\rho_x^{gal} \approx 0.3$ Gev/cm$^3$ $\approx 5 \times 10^{-25}$ g/cm$^3$ [2]. Experiments on search for dark matter are in progress in over 20 laboratories of the world. Most of them are aimed at direct registration of dark matter particles with high energy and mass. A specific technique is used in the experiment DAMA/NaI, DAMA/LIBRA which registers dependence on the time of a low background part of spectrum in the range of 2 - 6 keV [7]. The criterion of a dark matter signal is availability of annual background variations. Such a signal is detected in the experiment that has been conducted for ten years. Validity of annual variations with maximum on the 2-nd of June makes up 8.9 standard deviations [7]. It is this type of signal that should be expected to be produced by the solar system movement through the Galaxy halo at velocity of 230 km/s and the movement of the Earth around the Sun at velocity of 30 km/s. It results in variations of flux of dark matter particles coming through a detector. Exposure in the DAMA experiment is $2.5 \times 10^5$ kg $\times 24$ hours, exceeding a few orders of value that in other experiments. Another specific feature of the DAMA detector is a low threshold of signal registration, being equal to 2 keV. Fig. 1 [7] shows an energy spectrum of signals altering annually. One can see that the detector counting enhances as the registered energy decreases. The lowest point is at threshold of 2 keV. The further trend of spectrum in the range below the threshold is unknown. Theoretically, one can assume further increase of counting rate of events in decreasing energy. Such an increase of counting rate may result from interaction of dark matter particles being in some sense the long range. In this case the interaction cross section at low recoil energy grows in correspondence with a well-known dependence: $d\sigma/dE_R \propto 1/\varepsilon_R^2 v_{DM}^2$ where $\varepsilon_R$ - recoil energy, $v_{DM}$ - the velocity of dark matter particles.

---

[*] e-mail: serebrov@pnpi.spb.ru

The idea of considering the long-range dark matter, the interaction with which is usually described by the Yukawa potential is not a new one. For instance, the work [8] describes existence of the force of a new type in the dark matter sector with the Compton wavelength $m_\phi^{-1} \geq 1$ Gev$^{-1}$. In this work authors assume that the interaction of dark matter particles with the ordinary substance is implemented through mixing the field of a massive messenger of the interaction with an ordinary electromagnetic field. The work [9] treats the problem of scattering of multicomponent WIMP-particles on ordinary matter nuclei. The interaction between the dark matter particles and the ordinary matter occurs by mixing fields of a gauge massive boson from the dark matter sector with an ordinary electromagnetic field. For instance in the paper [10] results of the DAMA experiment are interpreted in terms of the Rutherford scattering of ordinary matter atoms by mirror matter atoms through exchange of a massless photon oscillating from the mirror state into the ordinary one and back. In this case interaction of the mirror matter with the ordinary one also results from mixing the field of a mirror photon with the ordinary electromagnetic field. The work [11] assumes using the model of an axion that interacts both with the mirror matter as well as with the ordinary one. In analogy with the paper [12] one can show that such an axion results in emerging the interaction of the Yukawa type. According to the above mentioned work [11] one succeeded in obtaining the constraint on the axion mass 1 MeV which corresponds to the Compton wave length of the order of $10^{-11}$ cm.

In addition to considering the long range interaction arising due to the exchange of interacting particles with bosons, the Kaluza-Klein model regards existence of external $\delta$ dimensions as causing some deviation of the Newton law at short distances to be parametrized by the Yukawa interaction $V(r) = -G_N \frac{m_1 m_2}{r}\left[1 + \alpha e^{-r/\lambda}\right]$ [13]. In this model radius $R$ of compactified space of $\delta$ dimension is connected with the Plank mass - $M_P \equiv G_N/\sqrt{8\pi} \simeq 2.4 \times 10^{18}$ GeV and energy scale magnitude of manifestation of external dimension - $M_D$ by the following relation: $M_P^2 \equiv R^\delta M_D^{2+\delta}$, in case $\delta = 2$, when $\lambda = R$. The knowledge of magnitude $M_D$ is useful for estimating cosmology and astrophysics effects associated with external dimensions. For instance the paper [14] states that evaluating the contribution into outer-space gamma radiation by decay of the Kaluza-Klein relic gravitons one could obtain constrain $M_D > 110$ TeV, which corresponds to $\lambda < 5.1 \times 10^{-6}$ cm.

There are different methods of search for long range forces in interactions of elementary particles, for instance, reference is made to the reviews [15]. In the field of short range distances ($10^{-11}$-$10^{-9}$ см) the investigation is performed using neutrons at energy of the order of electron volt [16]. In case of long range distances ($10^{-3}$-$10^{-2}$ cm) laboratory experiments are made on search for deviation from the interaction law $r^{-2}$ [17, 18]. In the area $10^{-8}$-$10^{-4}$ cm thermal and cold neutrons are used [19]. The works [20] suggest using data on measuring storage time of UCN in the trap with noble gases in order to analyze the contribution of long-range forces into neutron scattering. Besides it in the Ref. [21] constraints on value of CP violating forces between nucleons at distances of $10^{-4}$ – 1 cm were obtained.

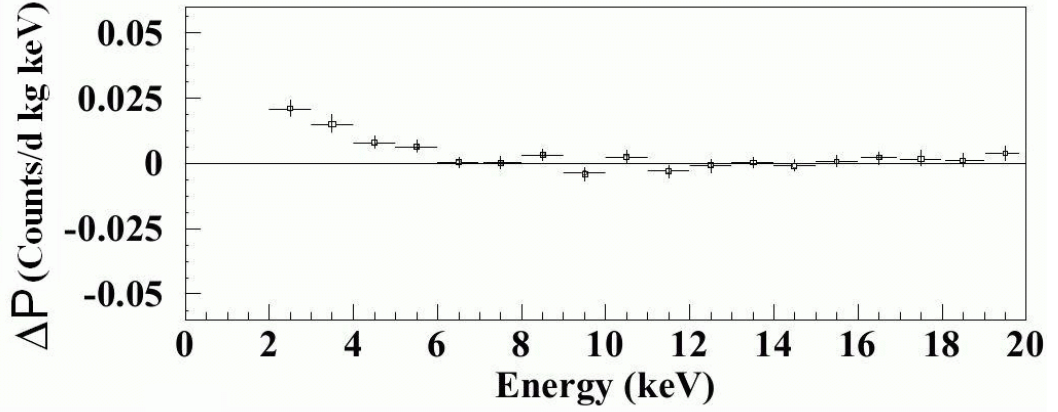

FIG. 1: Variation part of energy spectrum of signals with the annual period in the experiment DAMA

In accordance with the above made assumptions the question arises which will be under discussion in the following section. In all further considerations we will regard the signal observed in the experiment DAMA as being associated with dark matter in spite of the fact that such an assumption has not been proved yet.

## II. COULD AN ULTRACOLD NEUTRON TRAP BE TREATED AS A DETECTOR OF DARK MATTER PARTICLES WITH LONG-RANGE FORCES?

At first sight the answer to this question is negative, since the ultracold neutron density in a trap is by 20 orders of value less than the atom density in a substance. However, the registration threshold of dark matter with such a detector is by 10 orders of value less than that in the DAMA experiment. The matter is that an ultracold neutron having received the recoil energy $10^{-7}$ eV immediately escapes the trap, with this event being detected.

Now let us evaluate possible cross-section of elastic interaction of UCN and dark matter particles possessing long-range forces (LRF). We take the interaction potential of Yukawa form in the following parametrization:

$$U(r) = -\frac{\alpha G_N m_p m_x}{r} e^{-r/\lambda} \qquad (1)$$

Here $m_p$ - is the mass of the ordinary particle, $m_x$ - mass of Dark matter particle, $\lambda$ - is the length of the interaction and $\alpha$ - is a parameter which shows the ratio of forces in the gravitational and new interaction.

The Born approximation for integral cross-section of interaction with energy transfer grater than $\varepsilon_R^{min}$ and lower than $\varepsilon_R^{max}$ is expressed in the following way:

$$\sigma(\varepsilon_R^{max}, \varepsilon_R^{min}) = \frac{4\pi G_N^2 \alpha^2}{\hbar^4} \frac{(\varepsilon_R^{max} - \varepsilon_R^{min})}{E_x} \frac{\lambda^4 (m_p m_x)^3}{(2m_p \varepsilon_R^{max} \lambda^2/\hbar^2 + 1)(2m_p \varepsilon_R^{min} \lambda^2/\hbar^2 + 1)}, \qquad (2)$$

where $E_x$ is the energy of Dark matter incident particle. The value of $\varepsilon_R^{min}$ determined by the threshold energy of the detector and $\varepsilon_R^{max}$ can be calculated as $\varepsilon_R^{max} = 4 m_p m_x E_x / (m_p + m_x)^2$.

In terms of the DAMA analysis [22] one can assume the range of $\alpha$ values to be determined within $10^{26}$ - $10^{28}$. Since the registration of events in the DAMA experiment is impossible below the threshold of 2 keV, parameter $\lambda$ is not determined at all when we analyzed the DAMA signal. Consequently, the probability of long range character of interaction between dark matter particles with an ordinary matter does remain. If it does exist, it results in $\alpha \approx 10^{26}$. However, the range of $\lambda$ values is experimentally not determined because of rather high threshold of 2 keV. The value of $\lambda$ could be obtained, if there is a hypothetical possibility of reducing an experimental threshold. In this case differential cross-section (or the rate of counting events in Fig.1) will enhance with lowering $\varepsilon_R^{\min}$ until the following equation is satisfied: $2m_p \varepsilon_R^{\min} \lambda^2 / \hbar^2 \approx 1$. From this equation the value of the parameter $\lambda$ can be determined. Thus reducing the registration threshold one can determine the efficient radius of range of forces. UCN remain susceptible to contribution from the interaction (1) in the cross section (2) up to the magnitude $\lambda \simeq 10^{-6}$ cm.

### III. CONSTRAINTS ON $\alpha$ AND $\lambda$ PARAMETERS FROM UCN EXPERIMENTS

Now we will discuss what constraints on $\alpha$ and $\lambda$ could be already obtained from experiments with ultracold neutrons. We will consider experiments with neutrons in the ultracold trap.

#### A. Constraints from neutron decay time-life experiments

One can start this consideration with the roughest estimation making use of the following fact. In practice, in measuring neutron lifetime with UCN there may arise a systematic error resulting from UCN interaction with dark matter.

The probability of storing UCN in a trap is affected by a number of factors:
$$\tau_{st}^{-1} = \tau_n^{-1} + \tau_{loss}^{-1} + \tau_{vac}^{-1} + \tau_{DM}^{-1}.$$
Here $\tau_n^{-1}$ is the probability of neutron decay; $\tau_{loss}^{-1}$ - probability of losses caused by collisions with walls; $\tau_{vac}^{-1}$ probability of loses due to the residual vacuum and $\tau_{DM}^{-1}$ - probability of losses of neutrons from the trap due to collisions with dark matter particles.

The measurement of neutron lifetime on the beam in registering the decay fragments will not contain the last term contribution. Consequently, we are able to estimate the upper limit of the probability of UCN interaction with dark matter:

$$\left(\tau_n^{-1}\right)_{UCN} - \left(\tau_n^{-1}\right)_{beam} = \tau_{DM}^{-1} = (1.10 \pm 0.45) \times 10^{-5} \, s^{-1} < 2 \times 10^{-5} \, s^{-1}, \tag{3}$$

where $\left(\tau_n\right)_{UCN}$ - neutron lifetime measured by UCN technique (the most accurate result of measurements: 878.5±0.8 s [23]), $\left(\tau_n\right)_{beam}$ - neutron lifetime measured on neutron beam in registering the neutron decay fragments (the most accurate result of measurements: 886.8± 3.4 s [24]. Comparing the measurement results one can conclude that their difference does not exceed 2% at the confidence level 95%.

Now using the following equation: $n_x \sigma_{DM-UCN} v_{DM} \leq 2 \cdot 10^{-5} s^{-1}$, where $n_x$ - density of dark matter particles, $v_{DM}$ - velocity of dark matter particles, $\sigma_{DM-UCN} v_{DM}$ - integral cross section of elastic scattering of UCN and dark matter particles. We will use the formula (1) for calculating this cross section. The mean velocity of dark matter particles interacting with UCN depends on the component of DM concerned. If we consider the Galactic dark matter component, then this mean velocity is equal to 230 km/s (see below) but if we consider the dark matter gravitationally trapped by the Earth the velocity doesn't exceed the first cosmic speed of 8 km/s. According to the papers [25, 26] the dark matter density on the orbit of the Earth $\rho_x^E$ - is likely to exceed the Galaxy density of dark matter $\rho_x^{gal}$ -$10^3 - 10^4$ times. However, in a number of papers [27, 28, 29] it is stated that the density of captured dark matter is of the same order as the Galaxy density of dark matter. However, the above mentioned papers have not considered the long range interaction however. Below we will assume that due to the cross section increase at small energies, collision processes of dark matter in the substance of the gravitating object might give rise to the energy loss of the dark matter particle as well as to considerable coefficients of capture of dark matter into the gravitational well of the object. So far such calculations have not been made. We have made them for the Earth and are presenting here. To stress the influence of the captured dark matter one can introduce the corresponding capture coefficient. Constraints on this coefficient can also be obtained from UCN experiments. It is worth mentioning that it implies capture coefficient of low energy dark matter. Let us introduce capture coefficient of low energy dark matter. Capture coefficient of dark matter on the Earth orbit is as follows: $k_{capt}^{E} = \rho_{x}^{E}/\rho_{x}^{gal}$. Thus, $n_x$ is equal to: $n_x = (\rho_{x}^{gal}/m_x)k_{capt}^{E}$.

It is to be noted that capture coefficient is associated with energy of dark matter particles to be kept by the gravitational field of the Earth. For the DAMA experiment analysis the velocity of dark matter particles of 230 km/s has been used. It means that the Galaxy density has been included without any additional capture coefficient. Capture coefficient discussed above is related to the energy of dark matter particles, which is much lower than 2 keV threshold in the DAMA experiment. Yet these capture coefficient is closely concerned with the UCN experiment in which the registration threshold is very low. Let us see what constraints can be derived from the equation (2).

From the constraint $(\tau_n^{-1})_{UCN} - (\tau_n^{-1})_{beam} \leq 2 \times 10^{-5} s^{-1}$ one gets the constraint on value of $\alpha$ at big values of $\lambda$:

$$\alpha \leq 5.8 \times 10^{31} \sqrt{\frac{v_{DM}}{k_{capt}^{E} m_x}}. \qquad (4)$$

In this formula $v_{DM}$ is in m/s and $m_x$ is in nucleon mass. Fig.2 shows the area of constraints on parameters $\alpha$ and $\lambda$ (or $g_{aOM} g_{aDM}$ and $\lambda$) from Eq.(3) without taking into consideration the capture coefficient. The value of magnitude $\alpha$ from DAMA is calculated for big $\lambda$ at level $\approx 10^{26}$, or the value $g_{aOM} g_{aDM}$ at level $10^{-11}$. In Fig 2. the curve of constraints UCN-1 has been calculated with mass of dark matter particles of 20 nucleon units, i.e. equal to one of the allowed mass of the DAMA experiment. One can see that at radius of range of forces $10^{-4}$ cm constraints on parameters are by 6 orders weaker, than constraints on the values of $\alpha$ or $g_{aOM} g_{aDM}$ from DAMA. However, for the time being, these constraints do not take account of capture coefficient $k_{capt}^{E}$.

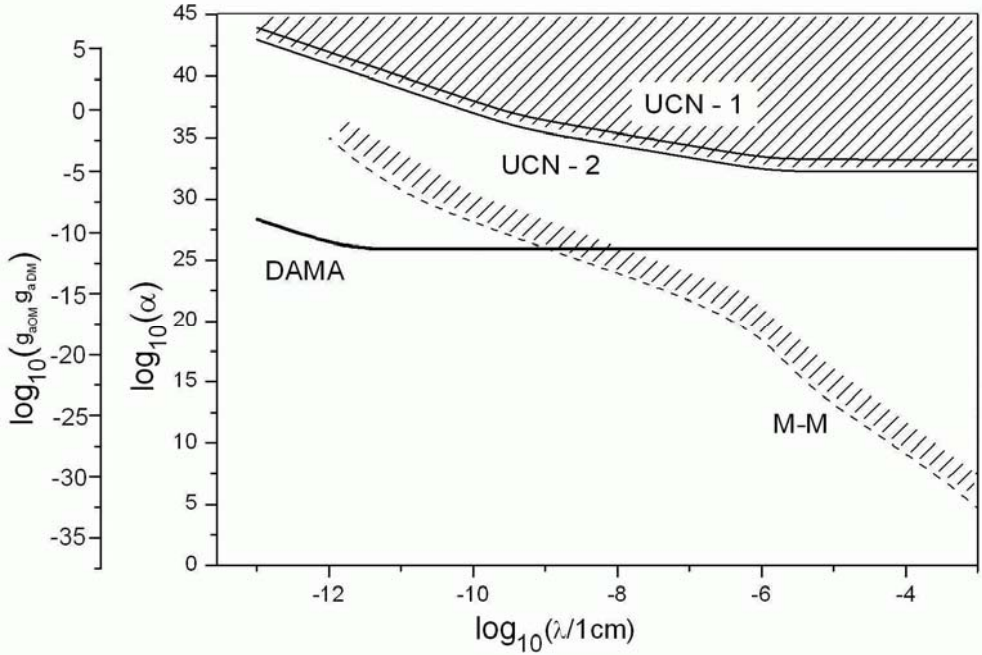

FIG. 2: Area of the parameters $\alpha$ and $\lambda$ (or $g_{aOM} g_{aDM}$ and $\lambda$) from DAMA experiment. The area of constraints UCN-1 is obtained from the equation (2). The area of constraints UCN-2 is obtained from the limit on probability of low-upscattering of UCN $(\tau_{up}^{-1}) < 3 \times 10^{-7} s^{-1}$ [30]. Area M-M corresponds to constraints on interactions of an ordinary matter with an ordinary matter [31].

One can make the maximal estimation of this coefficient, if it is supposed that the DAMA experiment did register the signal of the dark matter with long-range interaction $\lambda \approx 10^{-5} - 10^{-3}$ cm. Then, equating value $\alpha$ in formula (4) to our evaluation $\alpha = 10^{26}$ from DAMA, one can calculate the upper value of the coefficient $k_{capt}^E \leq 1.4 \times 10^{14}$. Correspondingly, the constraint on density of the long-range dark matter of the Earth is as follows $\rho_x^E < 6.6 \times 10^{-11}$ g/cm$^3$. The value $k_{capt}^E > 1.4 \times 10^{14}$ will run counter to the condition: $\left(\tau_n^{-1}\right)_{UCN} - \left(\tau_n^{-1}\right)_{beam} = \tau_{DM}^{-1} \leq 2 \times 10^{-5} s^{-1}$, i.e. in the experiment with UCN one would observe considerable suppression of neutron lifetime.

In Fig. 2 the area of M-M corresponds to the constraints on deviation from gravitational interaction at short distances for interactions of an ordinary matter with an ordinary matter [31]. In the area $\lambda = 10^{-12}$ cm the value $\alpha = 10^{26}$ from DAMA does not run counter to constraints M-M. If one transfers this value $\alpha$ for area $\lambda = 10^{-5}$ cm, then one turns out to be in the area which has been already excluded for interactions of an ordinary matter with an ordinary matter. Nevertheless we take interest in the interaction of dark matter with an ordinary matter. Therefore there remains possibility of discussing the theory of exchange of axion-like particles with different binding constants of this particle with ordinary and dark matter. From the equation of $g_{aOM}^2 \simeq 10^{-37} \alpha$ and experimental constraints $\alpha \leq 10^{15}$ at $\lambda = 10^{-6} - 10^{-5}$ cm it follows that $g_{aDM} \simeq 1$. For monopole-dipole interaction the experimental limits are much weaker $g_s g_p < 3 \times 10^{-12}$ at $\lambda = 10^{-5}$ cm [21]. In conclusion it should be noted that since there are practically no interaction models of dark and ordinary matter, the experimental constraints are of considerable value.

At first sight the possibility of using UCN trap for detecting dark matter with long-range forces on the Earth is not encouraging. However, it is to be taken into account that for long-range interaction the coefficient of capture of low energy dark matter by the Earth-Moon system has not been determined for the time being.

Maximal evaluation of amount of dark matter of the Earth was made in the paper [32], in terms of its possible influence on geophysical characteristics of the Earth. Maximal amount of dark matter on the Earth makes up $3.8 \times 10^{-3}$ from the mass of the Earth [32], i.e. the upper limit of the dark matter density on the Earth makes up $10^{-3} - 10^{-5}$ g/cm$^3$. Undoubtedly, it is extremely enormous and it would be unreasonable to use it in our estimations since from the maximal value of our estimation of the capture coefficient $k_{capt}^E$ it follows that $\rho_x^E < 6.6 \times 10^{-11}$ g/cm$^3$.

Now one should mention that the above evaluation of the sensitivity of an experiment with UCN trap to the search for dark matter is rather rough. The matter is that the evaluation is made by comparing the neutron lifetime, measured by different techniques and constitutes 1%. Here are used absolute measurements of neutron lifetime. If variations of UCN storage time are measured, the accuracy of measuring variations (relative measurements) may be 0.01%. In this case the experiment sensitivity to the determination of $\alpha$ may comprise: $\alpha_{sensitivity} \approx 6 \times 10^{30} \sqrt{v_{DM} / k_{capt}^E m_x}$. However, one should keep in mind that at small depth of variation the sensitivity of the method decreases. From this formula it follows that the experiment is rather sensitive to detecting big masses with long range of interaction. For instance for mass of dark matter particles $m_x = 10^{10}$ of nucleon mass $\alpha_{sensitivity} \approx 6 \times 10^{25} \sqrt{v_{DM} / k_{capt}^E}$. The UCN technique seems to allow super-heavy dark matter particles to be detected. Yet this conclusion is valid only in case the interaction constant is proportional to mass.

Finally, as far as the experimental task of detecting elastic scattering of dark matter particles with kinetic energy 5-10 eV is concerned, it is rather difficult to suggest the technique of detecting recoil energy below $10^{-7}$ eV. In this sense the UCN technique is unique.

**B. Experimental evaluation of constraints of the UCN heating effect in scattering on dark matter particles**

In the experiment [30] with UCN we studied thoroughly the effect of low-energy heating (the increase of kinetic energy) of UCN in reflecting from the trap walls. The heating effect was found to be $(2.2 \pm 0.2) \times 10^{-8}$ for one collision with the trap wall. One can use these data for estimating the upper limit on UCN heating due to interaction with dark matter particles. Frequency of UCN collisions in the trap in this experiment is 15 Hertz so the heating probability is less than $3 \times 10^{-7} s^{-1}$. To estimate the probability of heating of dark matter particles one can use the formula:

$$P_{up.DM} = n_x \cdot \sigma_{DM}^{up} \cdot v_{DM} < 3 \cdot 10^{-7} s^{-1} \qquad (5)$$

In this formula one must use the cross-section calculated with formula (1) using energy transfer from $\varepsilon_R^{min} = 0.5 \times 10^{-7}$ eV to $\varepsilon_R^{max} = 1 \times 10^{-7}$ eV, as it was in the experiment [30]. The result is as follows: $\sigma_{DM}^{up} = 6 \times 10^{-76} \alpha^2$ [cm$^2$]. In this calculation as it was above one uses $m_x = 20$. Thus from formula (4) one can set constraints on $\alpha$ at $\lambda = 10^{-5}$ cm: $\alpha < 2 \times 10^{32}$. The corresponding area of constraints on parameters $\alpha$ and $\lambda$ is shown in Fig. 2. It is approximately by one order of value better than the previous constraint from the measurements of neutron lifetime. These constraints have been also made without taking into account possible enhancement of dark matter density on the surface of the Earth. Correspondingly, evaluation on the maximal density of dark matter on the surface of the Earth is by two orders of value less than $\rho_x^E < 6.6 \times 10^{-13}$ g/cm$^3$.

**IV. DARK MATTER ACCUMULATION NEAR THE EARTH FOR LONG RANGE FORCES**

The dark matter (DM) in our Galaxy interacts with the Earth. The mean velocity of DM particles in the Galaxy is about 220 km/s. When such a particle is elastically scattered on the nuclei of the Earth it loses its energy and could be gravitationally absorbed by the Earth. As the escape velocity from the Earth is equal to 11.2 km/s the DM particle should be slowed down to lower velocity to be captured. After the DM particle is gravitationally absorbed it can stay at the stationary orbit around the Earth until it is captured by the Earth. Staying inside the Earth the DM particle will be thermalized and will get velocity distribution determined by the temperature of the Earth in the core. This temperature could be as high as 6000 K and the DM particle has noticeable probability to fly out of the Earth and in some cases even to escape the Earth if its velocity is higher than 11.2 km/s.

In this section we will calculate the rate of DM particles captured by the Earth, and also the lifetime of them in the Earth $t_{Earth}$ and inside the Earth gravitation $t_{life}$. Using these quantities we can calculate the additional density of DM particles on the Earth surface. As it follows in some cases this additional density will considerably exceed the DM mean density in our Galaxy.

**A. The Earth model**

The main properties of the Earth needed to describe the propagation of the DM in it are radial dependencies on the Earth density and the temperature, as well as on the Earth elemental composition. Radial dependence on the Earth density and temperature we used are presented in Figure 3. We suppose the Earth composition in the mantle and in the core to be different. Elemental composition of the Earth's mantle and core is presented in Table I [27]. The depth of the mantle is equal to 2900 km. In our calculation we assume that the mantle consists of the single element with average atomic mass equal to 23.5 and the core consists of iron only. We also divide the Earth volume into concentric layers with constant density and temperature obtained by averaging the real density and temperature distribution over the layer volume. The number of layers we considered was equal to 20.

TABLE I: The composition of the Earth's core and mantle

| Element | Mass number | Mass fraction Core | Mass fraction Mantle |
|---|---|---|---|
| Oxygen, O | 16 | 0.0 | 0.440 |
| Silicon, Si | 28 | 0.06 | 0.210 |
| Magnesium, Mg | 24 | 0.0 | 0.228 |
| Iron, Fe | 56 | 0.855 | 0.0626 |
| Calcium, Ca | 40 | 0.0 | 0.0253 |
| Phosphor, P | 31 | 0.002 | 0.00009 |
| Sodium, Na | 23 | 0.0 | 0.0027 |
| Sulfur, S | 32 | 0.019 | 0.00025 |
| Nickel, Ni | 59 | 0.052 | 0.00196 |
| Aluminum, Al | 27 | 0.0 | 0.0235 |
| Chromium, Cr | 52 | 0.009 | 0.0026 |

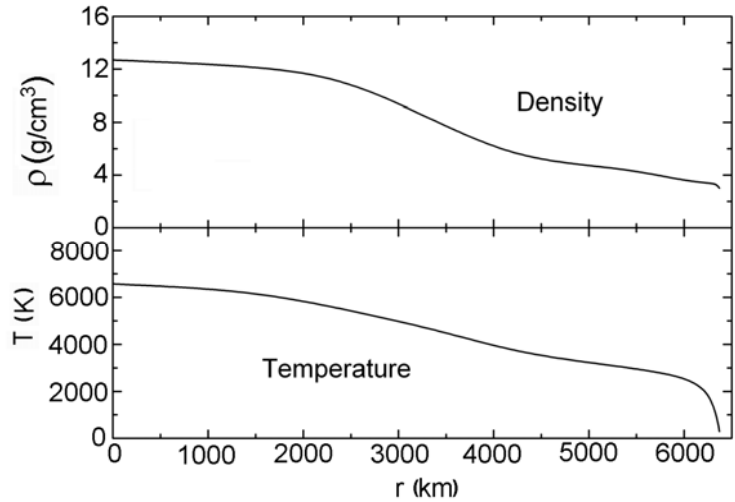

FIG. 3: Radial distribution of the Earth density and temperature

**B. DM propagation in the Earth**

We treat separately the influence of the gravitation and LRF on DM particle propagation in the Earth. To simplify simulation of DM propagation in the gravitational potential of the Earth we approximate potential to be constant in the layers into which we divide the Earth. As in region with constant potential the gravitational force is zero, the DM particle propagates freely inside the layers. The trajectory of particle breaks only at the border of layers. The particle can be either refracted or reflected from the border surface. The law of refraction has the following form

$$\begin{cases} v_1^2/2 + \phi_1 = v_2^2/2 + \phi_2 \\ v_1 \sin\theta_1 = v_2 \sin\theta_2 \end{cases}. \quad (6)$$

where $\phi_{1,2}$ are the gravitational potentials in regions 1 and 2, $v_{1,2}$ - velocity of particle in these regions, $\theta_{1,2}$ are the angles between the particle trajectory and the normal to the surface at the point where trajectory intersects the surface. In the case when equation system (6) has no solution the particle has mirror reflection from the surface.

Scattering on Yukawa potential of (1) can be treated in the Born approximation for interaction length $\lambda < 3 \times 10^{-2}$ cm. Scattering amplitude in the c.m. system has the following form

$$f(q) = \frac{2\mu}{\hbar^2} a \frac{\lambda^2}{(\lambda q)^2 + 1}, \quad (7)$$

where $\mu = m_x m_p/(m_x + m_p)$ is a reduced mass of scattering particles, $q$ is wave vector transfer, and parameter $a = -\alpha G_N m_x m_p$. Total cross section of scattering for different $\lambda$ and $m_x$ for $m_p$ = 20 GeV is presented in Figure 4. As is seen from the figure total cross section strongly depends on the interaction length and DM mass. For $\lambda > 10^{-6}$ cm interaction length becomes rather short and the simulation of DM propagation becomes very time consuming. Scattering angle in the c.m. system can be simulated according to the following formula

$$\cos\theta = \frac{1}{b}\left[b+1-\frac{1+2b}{1+2b\xi}\right], \quad (8)$$

where $\xi$ is a uniform random number distributed in the interval [0,1]. Parameter $b = 2\mu^2 v^2 \lambda^2/\hbar^2$ determines the asymmetry of scattering. The velocity distribution of the Earth particles is taken to be Maxwellian at the temperature of the layer where scattering occurs.

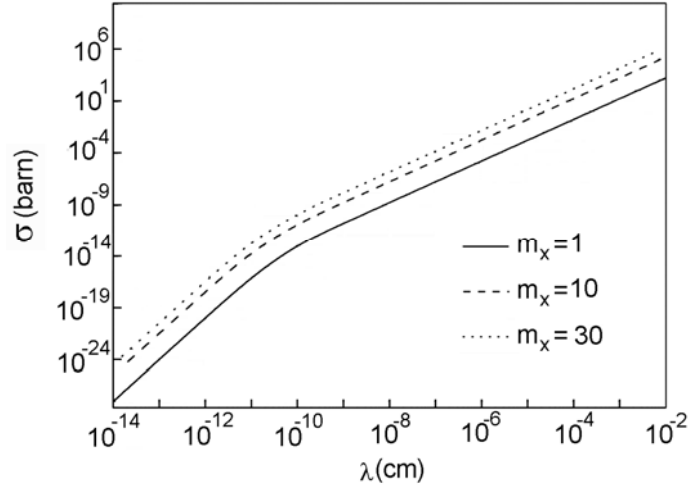

FIG. 4: Total scattering cross section of DM particle for different $\lambda$

## V. CAPTURE RATE OF DM PARTICLES BY THE EARTH

We take the velocity distribution of the Galaxy DM particles to be Maxwellian [33],

$$f_0(v)dv = n_x \frac{4}{\pi^{1/2}} \frac{v^2}{v_0^3} e^{-v^2/v_0^2} dv, \quad (9)$$

where the parameter $v_0$ = 220 km/s. We have taken the energy density of DM $\rho_x = n_x m_x$ = 0.3 Gev/cm$^3$. As the Sun with the Earth is moving in the Galaxy coordinate system at the velocity $v_S$ = 220 km/s, the velocity distribution of the DM particles from the Earth, as we see it, has the following form:

$$f(v) = n_x \frac{1}{\pi^{1/2}} \frac{v}{v_0 v_S}\left[e^{-(v-v_S)^2/v_0^2} - e^{-(v+v_S)^2/v_0^2}\right], \quad (10)$$

Velocity of DM particle increases when it approaches the Earth surface from the infinity according to the following equation: $v'^2 = v^2 + 2G_N M_E/R_E$. We assume the flux of DM particles to be isotropic if its asymmetry caused by the motion of the Sun is neglected.

After the DM particle falls on the Earth its propagation is considered according to the previous paragraph. During the elastic scattering on the Earth nuclei the DM particle loses its energy. We believe the DM particle to be captured by the Earth if it is reflected several times from the surfaces of layers into which the Earth volume was divided. This means that the particle energy was lowered enough to be trapped into the gravitational well of the Earth. After being trapped the particle is still living inside the Earth until it gets enough energy in the elastic scattering with the Earth matter to escape from it. Here we are not treating the interaction and probably annihilation of the trapped dark matter. Knowing the initial flux of DM and calculating in this manner the probability of capturing we have calculated the capture rate of DM by the Earth.

### A. Model testing

To test our model we have tried to calculate the test case of capture of weak interaction massive particles (WIMP) by the Earth. We suppose that these particles have fixed scattering cross section $\sigma_0$ with the nuclei of the Earth, and that this scattering is elastic and isotropic. We took this cross section to be equal to $\sigma = 10^{-34}$ cm$^2$. In paper [34] the capture rate of WIMP's by the Earth was theoretically calculated. In this test calculation we neglect the motion of the Solar system relative to the Galaxy coordinate system when the analytical formulas for the capture rate are the most transparent. When the Sun motion is taken into account the caption rate changes only by a factor not higher than 2.5. Also analytical formula doesn't take into account non zero temperature in the Earth. As was shown in [34] the influence of non zero Earth temperature on the capture rate is not crucial. Every element of the Earth gives its own contribution to the caption rate. Consider element of mass $m$ with number density $n$. Let its mass fraction in the Earth be $f$. It is convenient to introduce the following notations:

$$\mu = \frac{m_x}{m}, \quad \mu_\pm = \frac{\mu \pm 1}{2},$$
$$A = \frac{u^2}{v_0^2} \frac{\mu}{\mu_-^2}$$

(11)

where $u$ - is escape velocity of DM particle from the considered point inside the Earth. Let also $\hat{\phi}$ be the gravitational potential in the Earth relative to this potential on the Earth surface. The value of $\hat{\phi}$ is about 1.6 in the Earth center. We denote with $v_{esc}$ the escape velocity of particle on the Earth surface. This value is about 11.2 km/s. Then the caption rate of the DM by the considered element can be calculated as [34]:

$$C = \left[\left(\frac{8}{3\pi}\right)^{1/2} \sigma n_x \bar{v}\right] \left[\frac{v_{esc}^2}{v_0^2} \langle \hat{\phi} \rangle\right] \frac{M_E}{m} \left[\left\langle \frac{\hat{\phi}}{\langle \hat{\phi} \rangle} \left(1 - \frac{1-e^{-A^2}}{A^2}\right) \right\rangle\right], \quad (12)$$

where $\bar{v} = \sqrt{3/2}\, v_0$, $M_E$ is the mass of the Earth and angle brackets indicate averaging over the mass of the Earth. The total capture rate for all elements can be calculated as $\sum_f f C_f$, where $C_f$ is the capture rate calculated according to equation (12). The DM mass dependence on the capture rate has a maximum at $m = m_x$ for an individual element. Theoretically the calculated capture rate of WIMP's according to these formulas for our model of the Earth is presented in Figure 5(a). Here the result of our Monte Carlo simulation is also presented in comparison. We see that mass dependence of the theoretically calculated capture rate has two maxima according to the two average elements which have been considered: one with mass $m = 23.5$ GeV in the mantle of the Earth and the second at $m = 56$ GeV in the core. Monte Carlo results qualitatively agree with a theoretical prediction. Its mass dependence also has two maxima but the first maximum is about 20% smaller than the theoretical prediction and the width of the theoretical peaks is about two times smaller than that in our model. Probably this discrepancy between curve forms can be a consequence of neglecting the temperature of the Earth in the theoretical formulae.

## B. Capture rate of long range DM by the Earth

We have calculated the mass dependence on the capture rate for a long range interaction of DM with ordinary matter for different values of $\lambda$. Results are presented in Figure 5(b). For a long range interaction DM the mass dependence on capture rate is smooth without peaks. This is due to the fact that scattering cross section for such matter is high and DM particle has a lot of interactions in the Earth with small transfer of energy during each interaction. Also the absolute value of capture rate for $\lambda > 10^{-10}$ cm is about by two orders of magnitude larger than for WIMP's with cross section $\sigma = 10^{-34}$ cm$^2$. Capture rate increases with parameter $\lambda$ and is even higher for larger $\lambda$ than it is shown in Figure 5(b).

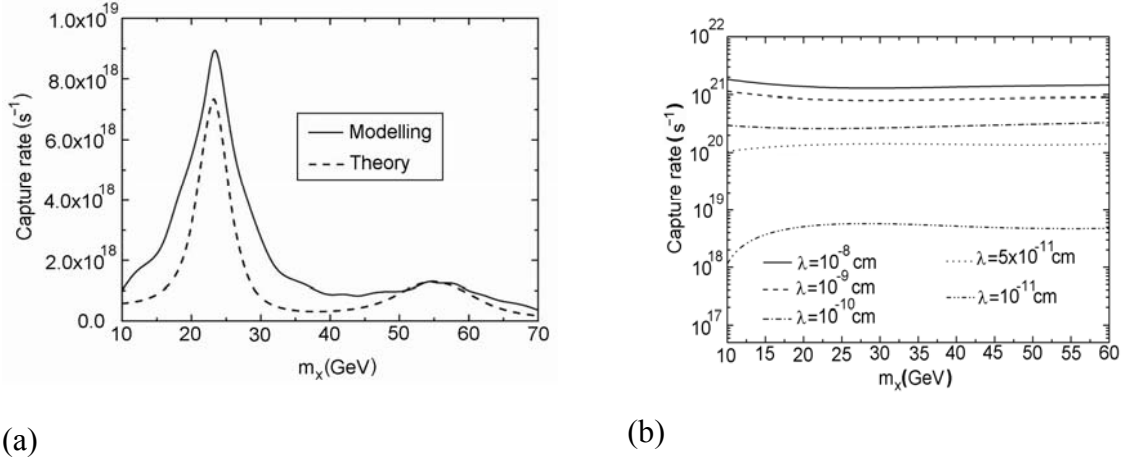

(a)  (b)

FIG. 5: Capture rate of WIMP's of different mass by the Earth

## C. DM evaporation

Captured DM thermolizes in the Earth so that its position-velocity distribution follows Maxwell Boltzmann law:

$$f_{th}(v,r) = \frac{n_0}{V_1} \frac{4}{\pi^{1/2}} \left(\frac{m_x}{2k_B T_x}\right)^{3/2} v^2 e^{-m_x v^2 / 2k_B T_x} e^{-m_x \Phi(r)/k_B T_x}, \qquad (13)$$

where $n_0$ - is the number of captured particles, $T_x$ - temperature of DM distribution, $\Phi(r)$ is the Earth gravitational potential, and

$$V_1 = \int_0^{R_E} 4\pi r^2 e^{-m_x \Phi(r)/k_B T_x} dr. \qquad (14)$$

is the effective volume of the Earth and $R_E$ is the Earth radius. Particles captured by the gravitational potential of the Earth can fly out of the Earth, move along the elliptic orbit around the Earth and then again return into the Earth. Sometimes if the velocity of the particle is larger than the escape velocity from the Earth the particle can fly out of the Earth. The latter process determines the total lifetime of the captured DM particle. The rate at which DM particle flies out of the Earth is determined by the mass of DM particle, difference of gravitational potential on the Earth surface compared with the center of the Earth and by the cross section with which DM particle interacted in the Earth. In Figure 6 (a) lifetime of DM particle in the Earth is presented for the particle of mass equal to 10 GeV and for different length of interaction $\lambda$.

Lifetime of DM particle in the Earth depends on the mass of DM particle $m_x$ according to the Boltzmann distribution:

$$t_{Earth}^{-1} \propto e^{-m_x \Delta\Phi(r)/k_B T_x}, \qquad (15)$$

where $\Delta\Phi$ is the difference of the gravitational potential between the center and the surface of the Earth. Calculated lifetime of DM particle in the Earth for different $m_x$ is compared with this distribution in Figure 6 (b). Calculated lifetime in Figure 6 (b) for small $m_x$ diverges from simple law (15) because the particles flying off the Earth have nonzero kinetic energies.

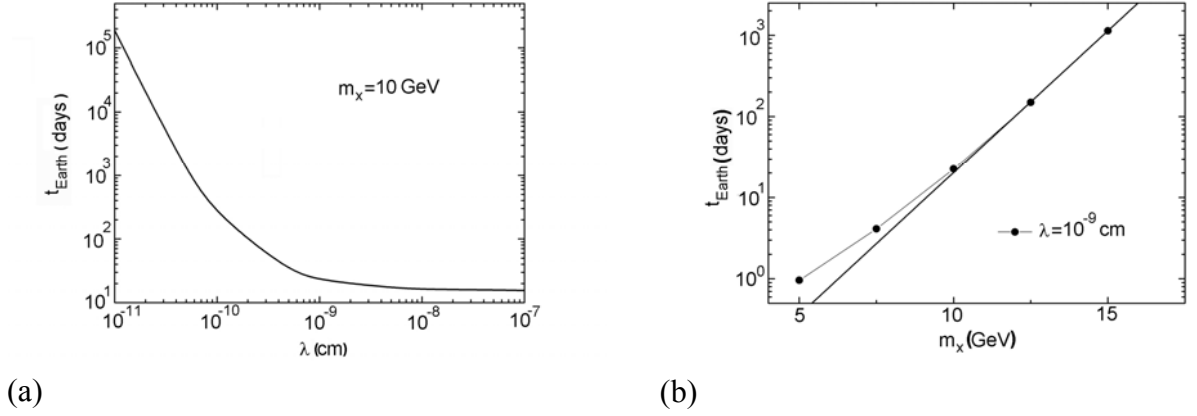

(a)  (b)

FIG. 6: Lifetime in the Earth for different $\lambda$ - (a); and for different $m_x$ - (b)

The total lifetime of the captured DM particle in the Earth gravitation is presented in Figure 7 for particle with mass $m_x = 10$ GeV for different $\lambda$. It has a peak for interaction length $\lambda$ equal to $10^{-9}$ cm. The total lifetime in the Earth gravitation is determined by the lifetime in the Earth $t_{Earth}$ which is divided by the probability of escaping the Earth. The latter probability depends on the form of the velocity spectrum of the particles which leave the Earth. Computer simulation shows that this spectrum for $\lambda > 10^{-9}$ cm can be described by simple Gaussian with nearly constant dispersion. On the other hand, the mean velocity of DM particle flying out of the Earth slightly rises with $\lambda$ here so that probability of escaping the Earth gravitation increases and the total lifetime in Fig. 7(a) falls for large $\lambda$.

Dependence of total lifetime of DM particle on $m_x$ follows the similar law (15) but with $\Delta\Phi = \Phi(\infty) - \Phi(0)$, where $\Phi(0)$ is the gravitational potential in the center of the Earth. Calculated total life time is compared with law (15) in Figure 7(b). As is seen from the comparison the total lifetime exactly follows Boltzmann law. Fitted temperature of the captured DM in the Earth is equal to 4900 K which is about 75% of the maximal temperature in the Earth core. Total lifetime of heavy DM particles in the gravitational potential of the Earth exceeds the age of the Earth. This time determines that of DM accumulation by the Earth. The accumulation time is equal to the total lifetime of the DM particle if it doesn't exceed the age of the Earth, or simply equals the age of the Earth if total lifetime is higher than the Earth age.

### D. DM density on the Earth surface

Captured by the Earth gravitation DM particles constantly fly out of the Earth, stay some time at elliptic orbit and then return to the Earth. As the age of the Earth is rather large (about 4.5 billion years) and the rate with which LRF DM is captured by the Earth is also large the amount of DM accumulated in the Earth may be up to $10^{37}$ particles. Accumulated by the Earth DM particles can produce additional DM density on the Earth surface. This density can be calculated as

$$\rho_x = m_x \frac{C t_{accum}}{(t_{Earth} + t_{orb}) S_E \bar{v}_{orb}}, \quad (16)$$

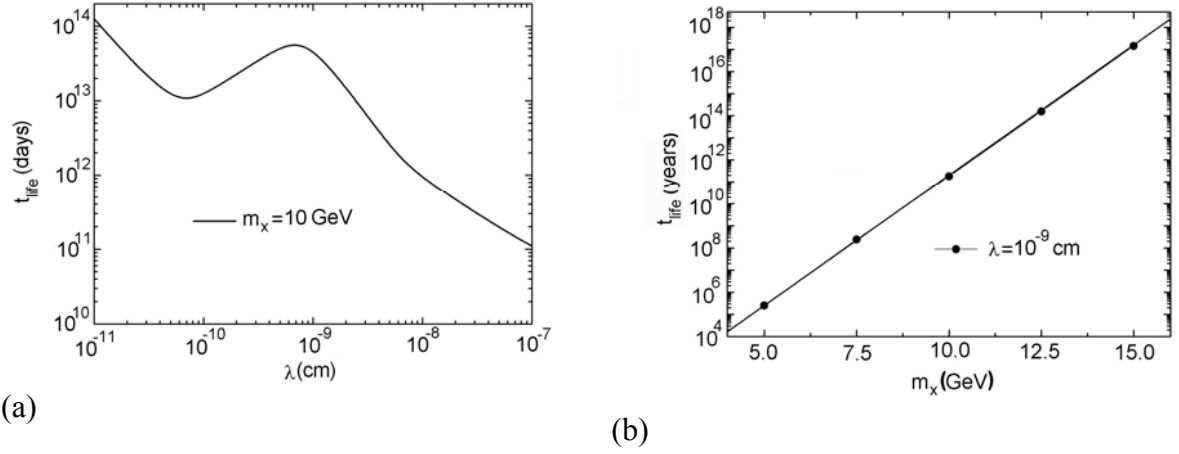

FIG. 7: Total lifetime of DM particle before escaping for different $\lambda$ - (a); $m_x$ - (b)

where $S_E$ is the area of the Earth surface, $t_{orb}$ is the average time the DM particles stay at the orbit outside the Earth which is usually not larger than 0.1 day, and $\bar{v}_{orb}$ is the mean DM particle velocity when it leaves the Earth surface. In Eq. (16) $C$ is the capture rate of DM particles by the Earth, and $t_{accum}$ is the accumulation time of DM by the Earth, which is equal to total lifetime of DM particle captured by the Earth gravitation if it doesn't exceed the Earth age. In the numerator of this equation we have the total amount of dark matter particles accumulated in the Earth gravitation well until they escape it. When we divide this number by the total time the particle stay captured inside the Earth $t_{Earth}$ and the time they stay on the orbit - $t_{orb}$ we get the rate at which the particle fly through the Earth surface. Dividing the rate by the square of the Earth surface we get the flux of the captured dark matter particles on the Earth surface. As usual the density of particles can be obtained from the flux dividing it on the mean velocity of particles. And here it is $\bar{v}_{orb}$ - the velocity at which the particles flying off the Earth are moving on the orbit. The results of the calculations of additional DM density for different interaction length $\lambda$ and DM particle mass $m_x$ are presented in Figure 8. As seen from this figure for DM particle masses smaller than 15 GeV the additional DM density could reach rather high values higher than $k_{capt}^{E} \simeq 10^6$ times exceeding the mean DM density in our Galaxy. From (4) it follows that constraints on $\alpha$ in this case will be about $10^3$ stronger.

However even with this factor, constraints on $\alpha$ from neutron lifetime measurements are still weaker than it is needed to exclude the case of dark matter interaction with ordinary matter by long range forces. At the same time, we can not take into account any additional mechanism and the value of $k_{capt}^{E}$ may be even larger. In this case we have a chance to register the signal from dark matter by measuring time variation of neutron storage in the neutron trap. A periodic character of the signal similar to that occurring in the DAMA experiment can be regarded to be a criterion of the dark matter signal in the experiment with UCN trap. However, variations of dark matter flux on the surface of the Earth, where the UCN trap is located, can be caused by reasons different from those in the DAMA experiment. The dark matter captured by the Earth can be involved by the Moon in such a way that the density of dark matter on the Earth surface will be changed depending on the hour of the day. The first experiment to measure such possible variation with the neutron trap will be treated in the next section. One can see that the dark matter density on the Earth surface at night is higher than that during the day time, i.e. one should expect 24-hours variations of dark matter density at the detector, but the phase of 24-hours variations is expected to change regularly due to rotation of the Moon around the Earth. For instance, the phase is likely to change by 180 degree in the position of full Moon with respect to the phase of new Moon. In fact the oscillation period of dark matter density will constitute 24 hours 50 minutes rather than 24 hours, since the Moon will appear over the detector exactly in correspondence with this period.

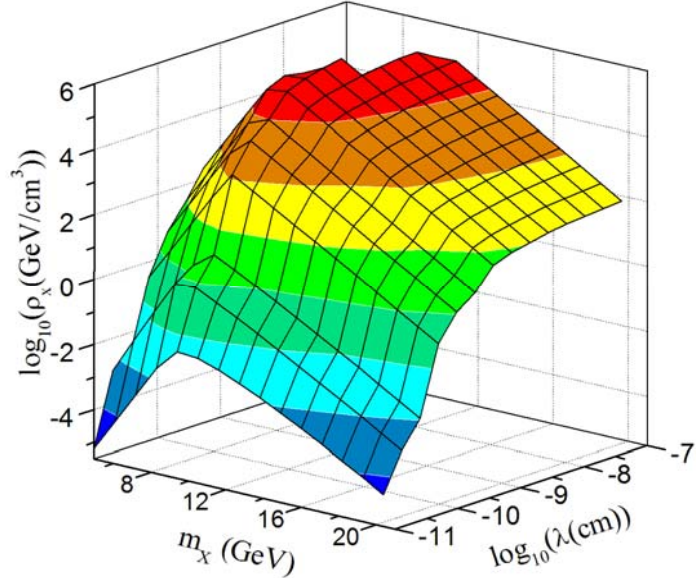

FIG. 8: Additional DM density on the Earth surface

## VI. THE FIRST EXPERIMENTS ON OBSERVATION OF TIME VARIATION OF NEUTRON STORAGE

Measurements of time variation of UCN storage in a trap have been made in the course of the experiment on search for neutron - mirror neutron oscillations [35]. This experiment is known to be aimed at finding correlation of time storage with the value of magnetic field in UCN trap. Such a correlation with accuracy of $\Delta\tau/\tau \leq 10^{-4}$ has not been detected. These measurement data can be used in order to evaluate possible time variations of UCN storage in a trap. The experimental diagram is shown in Fig. 9(a). The trap is filled with UCN through the central neutronguide at the open inlet valve 8 and at the closed outlet valves 7 and 9. Then the valve 8 is closed. After holding UCN in a trap during $t_{hold}$ valves 7 and 9 are opened and UCN remaining in the trap are registered by detectors 5 and 6.

The process of trap filling (1) and the process of holding (2) can be observed as valves 7, 9 have not been completely closed. Two cases with time of holding 50 s. and 470 s. are shown.

Fig. 9(a) shows the installation operation 9(b) measuring cycles: 1 - trap filling, 2 - UCN storage, 3 - UCN release on detectors, 4 - measurement of detector background. Being measured, the UCN density has been controlled during the process of filling the trap with monitoring detector 12 installed on the input neutron guide. These measurements enabled to take into consideration fluctuations of UCN density in the source of UCN. For the time storage variation to be determined one measured the quantity of UCN $N(300)$ registered by detectors 5, 6 after a storage cycle during $t_{hold}$, equal to 300 s. The measured data were normalized into the recordings of the monitoring detector $N_M$, measured in filling the trap. Thus normalization of the effect by value $N_0$ could be done by monitoring recordings. This fact is of particular importance as such normalization takes account of fluctuations of UCN source just at the moment of filling the trap.

There is the storage time drift in the course of the experiment during the period from the 29th of August till the 22nd of September 2007 and from 10th of November till the 5th of December.

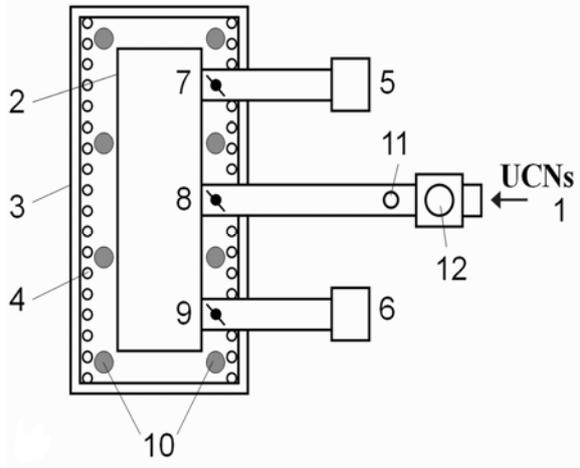 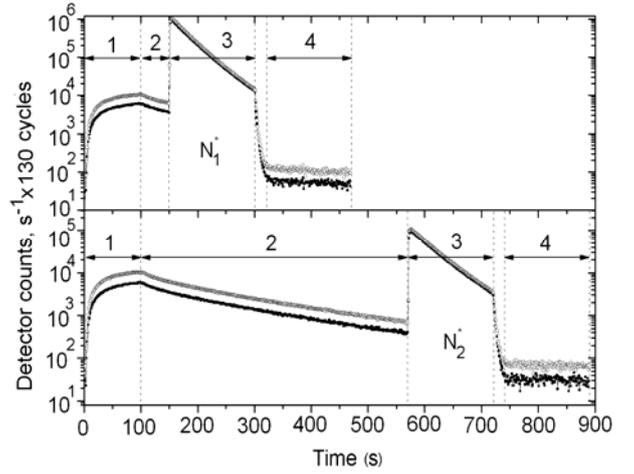

(a) (b)

FIG. 9: (a,b) (a) experimental scheme. 1- input neutronguide, 2 - UCN storage trap, 3 - magnetic shielding, 4 - solenoid, 5,6 - UCN detectors, 7,8,9 - UCN valves, 10 - magnetometers, 11 - monitoring detector, 12 - inlet valve, (b) rate of detector counting 5 and 6 in the course of measurements.

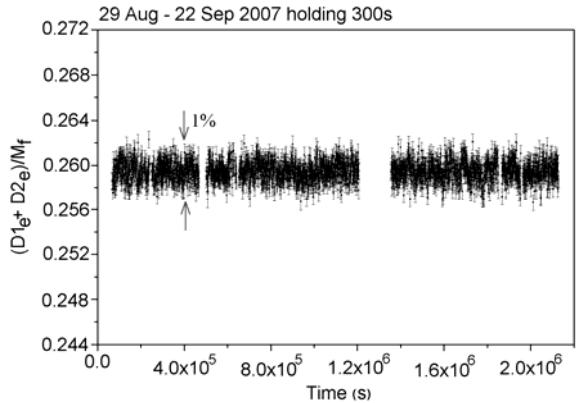 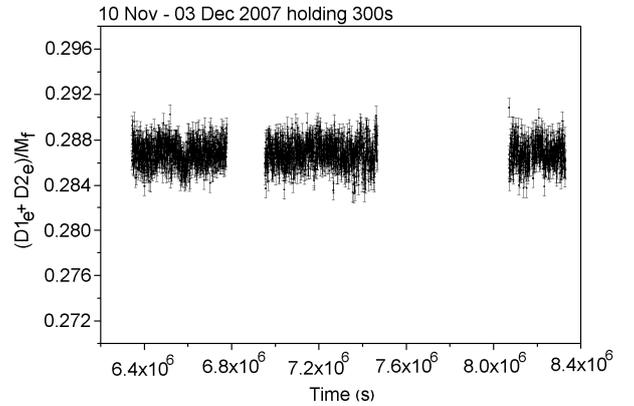

(a) (b)

FIG. 10: Behavior of summarized recordings of detectors 5, 6 after UCN storage in a trap for 300 s. in the course of measurements (a) 29August - 22 September 2007 and (b) 10 November - 03 December 2007 after correcting on the drift of time storage of UCN in the trap

Owing to degasation of the trap surface in pumping the storage time smoothly increases. In general it results from removing the residual water from the trap surface. Fig. 10 shows the summarized recordings of detectors 5, 6, normalized to the count rate on the monitor detector (11) after UCN storage in a trap for 300 s. Correcting the drift enables to obtain regular recordings with statistical variations. It is impossible to detect any daily variations because of big statistical errors. Yet, of special interest to us are periodic oscillations either within 24 hours or within 24 hours 50 minutes. Summing up measurements in these time periods we are capable of distinguishing a periodic signal and of averaging casual fluctuations.

Fig. 11 presents measurement processing with the period of 24 hours, while Fig. 12 presents measurement processing with the period of 24 hours 50 minutes. In the latter case the time scale of measurements was recalibrated according to lunar hours. All lunar periods of 24 hours were overlapped. As a result, every hour got a great number of measurements, with statistical accuracy of measurements being increased.

One has failed to detect any variations with the accuracy about $10^{-4}$. By combining the data, collected during the period 29 August-22 September 2007, and data, collected during the period 10 November - 03 December 2007, the restrictions on time variations of storage with period 24 hours and 24 hours 50 minutes were obtained $(-0.64 \pm 0.84) \times 10^{-4}$ and $(-1.7 \pm 0.8) \times 10^{-4}$, respectively. The upper constraint (95%C.L.) on time variations of storage with period of 24 hours is equal to $2.0 \times 10^{-4}$, while that with period of 24 hours 50 minutes it is equal to $3.5 \times 10^{-4}$.

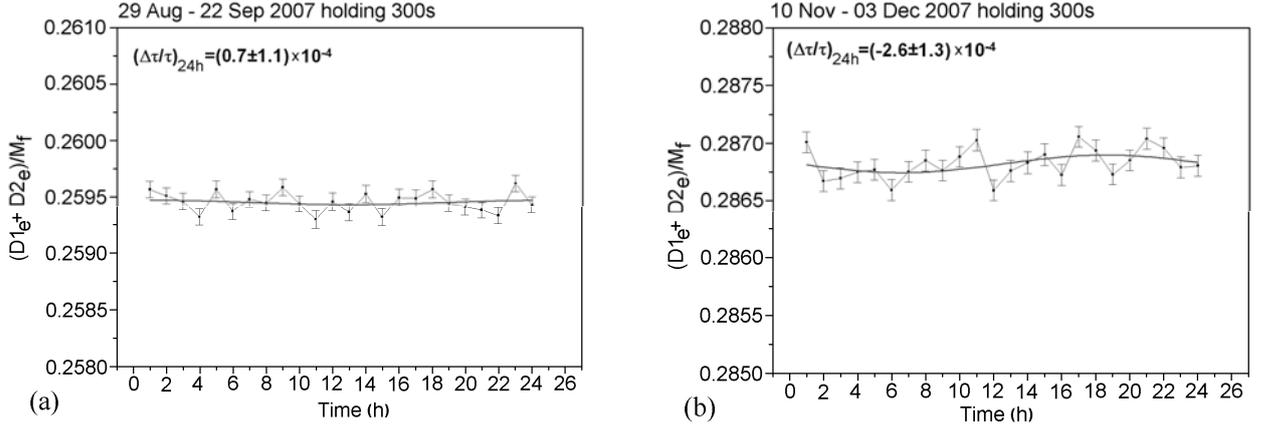

FIG. 11: Data treatment on variations with period of 24-hours. (a) Data were collected during the period 29 August – 22 September 2007, (b) Data were collected during the period 10 November - 03 December 2007

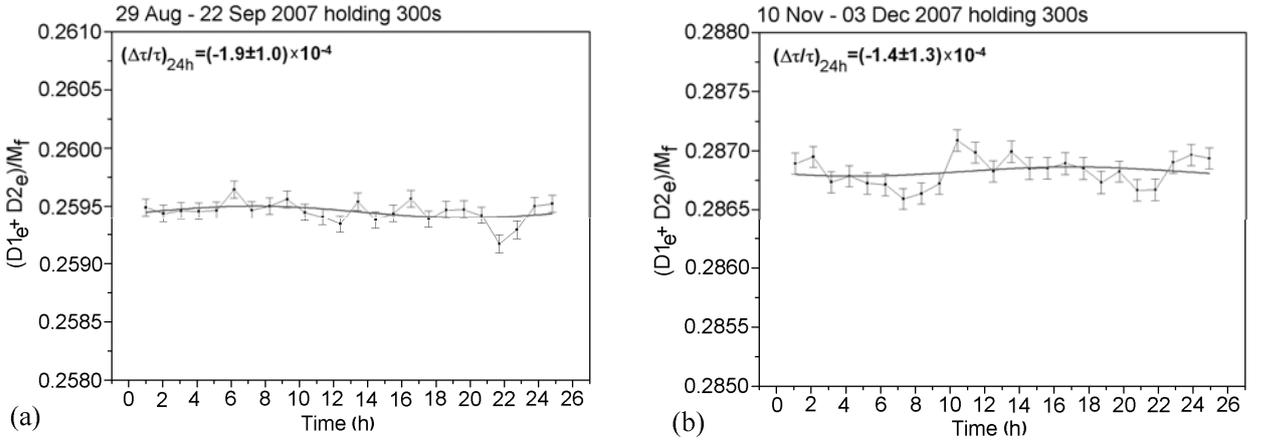

FIG. 12: Data treatment on variations with period of 24-hours 50 minutes. (a) Data were collected during the period 29 August - 22 September 2007, (b) Data were collected during the period 10 November - 03 December 2007

The storage time of UCN in a trap is 190 s. If these upper constraints are recalculated with respect to variations of neutron lifetime, the corresponding upper limits 0.5 s and 0.9 s will be obtained.

## VII. CONCLUSIONS

The main assumption of the present paper is concerned with introducing the long range interaction ($\lambda$ order of $10^{-10} - 10^{-5}$ cm) for dark matter particles. Consequently the following conclusions can be drawn. Of considerable interest is the assumption on complementary interaction (in addition to gravitational character) of dark matter particles with those of ordinary matter at the distance about $\lambda$. The first experimental constraints are of great significance. Probability of dark matter impact on UCN storage probability does not exceed $2 \times 10^{-5} s^{-1}$ as compared with data on neutron life time $\left(\tau_n^{-1}\right)_{UCN} - \left(\tau_n^{-1}\right)_{beam} = \tau_{DM}^{-1} < 2 \times 10^{-5} s^{-1}$. Probability of dark matter impact on heating UCN does not exceed $3 \times 10^{-7} s^{-1}$ as follows from the direct experiment on low energy heating UCN [30]. If one assumes that in the DAMA experiment the long-range dark matter signal is detected, the UCN experimental data provide us with opportunity to set a constraint on the density of the long-range dark matter on the Earth: $\rho_x^E < 6.6 \times 10^{-11}$ g/cm$^3$.

Upper limits (95%C.L.) are set on variations of neutron life time with period of 24 hours and with that of 24 hours 50 minutes equal to 0.5 s and 0.9 s correspondingly.

To summarize, one should say that preliminary evaluations of sensitivity of "DM-UCN trap" experiment with respect to that of DAMA are not encouraging due to low density of UCN.

However, use of long-range interaction gives rise to possibility of capturing and accumulating low-energy dark matter in the gravitational field of the Earth. In this case the UCN technique is of great practical value as recoil energy at level of fractions of electron volt can be detected only in the experiment applying the UCN trap. Carrying out the proposed UCN technique seems to be highly expedient as it gives way to a new field of exploring the low energy matter with long range forces.

## VIII. ACKNOWLEDGMENTS

The authors would like to express their gratitude to Z. Berezhiani, A.D. Dolgov, E.M. Drobyshevski, D.S. Gorbunov, M.Yu. Khlopov, V. Kauts, M.S. Lasakov, V.A. Rubakov and D.A. Varshalovich for useful discussions and critical remarks. The idea of this experimental technique arose during the visit to Z. Berezhiani at the Grand Sacco National Laboratory in February 2007. This investigation was supported by the Russian Foundation for basic research (projects:08-02-01052a, 10-02-00217a, 10-02-00224a, 11-02-01435-a, 11-02-91000-ANF-a) and by the Federal Education Agency (contracts: P2427, P2540), as well as by the Federal Agency of Science and Innovations (contract 02.740.11.0532) and the Ministry of Education and Science of the Russian Federation (contract 14.740.11.0083)